\documentclass[prl,a4paper,aps,twocolumn,
               nofootinbib,nobibnotes,superscriptaddress]
              {revtex4}
              
\pdfoutput=1

\usepackage[dvips]{graphicx}
\usepackage{amsmath,amssymb,mathrsfs}
\usepackage{bm}
\usepackage{times}
\usepackage{epsfig}
\usepackage{verbatim}
\usepackage{bm}
\usepackage[utf8]{inputenc}
\usepackage{graphics}
\usepackage{graphicx,epsfig,amssymb,amsmath,color,cancel}
\usepackage{subfigure}
\usepackage[english]{babel}
\usepackage{slashed}
\usepackage{chngcntr}
\usepackage{appendix}

\newcommand{\be}{\begin{equation}}
\newcommand{\ee}{\end{equation}}
\newcommand{\ba}{\begin{array}}
\newcommand{\ea}{\end{array}}
\newcommand{\bea}{\begin{eqnarray}}
\newcommand{\eea}{\end{eqnarray}}

\begin{document}

\title{Primordial Black Holes from the QCD Axion}

\author{Francesc Ferrer}
\affiliation{Department of Physics, McDonnell Center for the Space Sciences,
Washington University, St. Louis, Missouri 63130, USA}
\affiliation{IFAE and BIST, Universitat Autònoma de Barcelona, E-08193 Bellaterra, Barcelona}
\author{Eduard Masso}
\affiliation{IFAE and BIST, Universitat Autònoma de Barcelona, E-08193 Bellaterra, Barcelona}
\affiliation{Departament de F\'{\i}sica, Universitat Autònoma de Barcelona, E-08193 Bellaterra, Barcelona}
\author{Giuliano Panico}
\affiliation{IFAE and BIST, Universitat Autònoma de Barcelona, E-08193 Bellaterra, Barcelona}
\affiliation{Laboratory of High Energy and Computational Physics, National Institute of Chemical Physics and Biophysics, Rävala pst. 10, 10143 Tallinn, Estonia}
\author{Oriol Pujolas}
\author{Fabrizio Rompineve}
\affiliation{IFAE and BIST, Universitat Autònoma de Barcelona, E-08193 Bellaterra, Barcelona}

\date{\today}

\begin{abstract}
\noindent We propose a mechanism to generate Primordial Black Holes (PBHs) which is independent of cosmological inflation and occurs slightly below the QCD phase transition.~Our setup relies on the collapse of long-lived string-domain wall networks and is naturally realized in QCD axion models with domain wall number $N_{DW}>1$ and Peccei-Quinn symmetry broken after inflation. In our framework, dark matter is mostly composed of axions in the meV mass range along with a small fraction, $\Omega_{\text{PBH}} \gtrsim 10^{-6} \Omega_{\text{CDM}} $ of heavy $M \sim 10^4-10^7 M_\odot$ PBHs. The latter could play a role in alleviating some of the shortcomings of the $\Lambda$CDM model on sub-galactic scales. The scenario might have distinct signatures in ongoing axion searches as well as gravitational wave observatories.
\end{abstract}

\maketitle

\paragraph{\textbf{Introduction.}}

The recent detection of gravitational waves emitted by the merging of relatively heavy black holes ($M\gtrsim O(10) M_{\odot}$) \cite{Abbott:2016blz} has revived interest in the proposal that the DM of the universe comprises Primordial Black Holes (PBHs)~\cite{Hawking:1971ei, Carr:1974nx, Bird:2016dcv, Clesse:2016vqa, Sasaki:2016jop, Kashlinsky:2016sdv}. Although there are constraints on the abundance of PBHs for almost all viable masses (see e.g.~\cite{Carr:2016drx}), a small relic abundance of heavy ($M\gtrsim 10^{5} M_{\odot}$) PBHs may play an important role in the
generation of cosmological structures and alleviate shortcomings of the CDM scenario on sub-galactic scales~\cite{Clesse:2015wea,Carr:2018rid}. Furthermore, such PBHs could shed light on the origin of the super-massive BHs in the centers of most galaxies, some of which were already in place at very early times~\cite{Wu:2015,Banados:2017unc}. 

Several fundamental physics scenarios may explain the existence of PBHs. Arguably, the most studied proposal relies on the gravitational collapse of density fluctuations generated during inflation (see e.g.~\cite{Carr:2016drx}). Nevertheless, it is interesting to understand whether PBHs could naturally arise in other contexts.

In this Letter we propose an alternative PBH formation mechanism, independent of inflationary physics, that relies on the collapse of axionic
topological defects (see e.g.~\cite{Vilenkin:2000jqa} for an introduction). The generation of PBHs from defects has been investigated in different contexts including PBHs from the collapse of string loops~\cite{Vilenkin:1981iu, Hawking:1990tx, Fort:1993zb, Garriga:1993gj}, and from domain walls (DWs) during inflation~\cite{Khlopov:2004sc, Deng:2016vzb}. Here we discuss for the first time the formation of PBHs from long-lived string-DW networks~\cite{Kibble:1982dd, Vilenkin:1982ks}~(see~\cite{Vachaspati:2017hjw} for a setup closely related to ours and~\cite{Sakharov:1994id, Sakharov:1996xg, Khlopov:1999tm} for similar previous work) appearing in well-known realizations~\cite{Kim:1979if, Shifman:1979if,Dine:1981rt, Zhitnitsky:1980tq} of the Peccei-Quinn (PQ) solution to the strong CP 
problem~\cite{Peccei:1977hh, Wilczek:1977pj, Weinberg:1977ma}. These so-called hybrid networks have multiple DWs attached to strings\footnote{This situation can arise more generally from sequences of phase transitions in the early Universe.}, and suffer from a DW problem~\cite{Zeldovich:1974uw} unless the vacua separating different walls are split~\cite{Sikivie:1982qv}. The splitting likely requires extra new physics beyond the QCD axion (see below), however this need not interfere with the present mechanism of PBH formation.

When the PQ symmetry is broken after inflation, the axion abundance receives comparable contributions from the 1) misalignment 
mechanism, 2) radiation from string defects~\cite{Klaer:2017ond, Gorghetto:2018myk, Kawasaki:2018bzv}, and 3) annihilation of the string-wall network~\cite{Kawasaki:2014sqa} (see also~\cite{Ringwald:2018xlf}). We show in this Letter that there can be a fourth small contribution to the axion DM abundance in the form of heavy, $10^{4-7} M_{\odot}$, PBHs. Interestingly, this provides a concrete realization of the proposed role of massive PBHs in the early Universe~\cite{Carr:2018rid} in the context of QCD axion DM. Moreover, our scenario is not subject to some of the strong constraints arising from CMB $\mu$-distortions, which plague PBH formation mechanisms from gaussian inflationary fluctuations (see~\cite{Carr:2018rid}).

The hybrid network dynamics is hard to analyze. However, for our purposes the essential features can be captured by focusing on the closed walls that arise in the network~\cite{Vachaspati:2017hjw}.

\paragraph{\textbf{Collapse of closed domain walls.}}

Once the Hubble length becomes comparable to the closed wall size $R_{\star}$, the DW rapidly shrinks because of its own tension. This occurs at the temperature $T_{\star}$ defined by $R_{\star}\sim H^{-1}_{\star}\simeq g_{\text{eff}}(T_{\star})^{-1/2} M_{p}/T_{\star}^{2}$, where $M_{p}=(8\pi G_{N})^{-1/2}$ and $g_{\text{eff}}(T_{\star})$ is the effective number of degrees of freedom at $T_{\star}$. 
The total collapsing mass has two contributions: one induced by the wall tension $\sigma$, and another one coming from any possible difference in energy density between the two regions separated by the DW:
\begin{equation}
\label{eq:mass2}
M_{\star}=4\pi\sigma R_{\star}^2+\frac{4}{3}\pi\Delta\rho R_{\star}^{3}\sim 4\pi\sigma H^{-2}_{\star}+\frac{4}{3}\pi\Delta\rho~ H^{-3}_{\star}.
\end{equation} 
For closed DWs arising in the network $\Delta\rho\geq 0$~(see below), thus the wall bounds a region of false vacuum.

Another important parameter for the formation of PBHs is the ratio of the Schwarzschild radius $R_{S}$ of the collapsing wall to the initial 
size $R_{\star}$ :
\begin{equation}
\label{eq:merit2}
p\equiv \frac{R_{S}}{R_{\star}}\sim \frac{2G_{N} M_{\star}}{H^{-1}_{\star}}\sim \frac{\sigma H^{-1}_{\star}}{M_{p}^2}+\frac{\Delta\rho~ H^{-2}_{\star}}{3 M_{p}^{2}}.
\end{equation}
If $p$ is close to $1$ then the DW rapidly enters its Schwarzschild radius and forms a BH. If $p\ll 1$, however, the wall has to contract 
significantly before falling inside $R_{S}$. It is then less likely to form a BH, since asphericities, energy losses and/or angular momentum may severely affect the dynamics of the collapse. We will thus refer to $p$ as the \emph{figure of merit} for PBH formation from the collapse of DWs. 

The temperature behavior of $p$ and $M_{\star}$ is crucial to our proposal. Whenever the tension terms dominate in~\eqref{eq:mass2} and~\eqref{eq:merit2}, we have $M_{\star}\sim T_{\star}^{-4}$, $p \sim T_{\star}^{-2}$. If, instead, the energy difference terms dominate, we have $M_{\star}\sim T_{\star}^{-6}$, $p \sim T_{\star}^{-4}$. 

Therefore, the duration of the hybrid network has a huge impact on the likelihood of forming PBHs, as well as on their masses. The use of long-lived string-wall networks is the essential new idea of our proposal. This requires multiple DWs attached to each string~\cite{Vilenkin:2000jqa}. Interestingly, this can be realized in QCD axion models with domain wall number larger than one.~\footnote{The original DFSZ \cite{Dine:1981rt, Zhitnitsky:1980tq} axion has $N_{DW}=6$, while the simplest KSVZ realization \cite{Kim:1979if, Shifman:1979if} has $N_{DW}=1$. However, generalizations of the latter with $N_{DW}>1$ can be considered.} 

\paragraph{\textbf{Axion Dark Matter from String-Wall Networks.}}

Let us embed the basic mechanism illustrated in the previous section in the QCD axion cosmology. Consider a scalar field $\Phi$ with a $U(1)_{PQ}$ symmetry broken at some temperature $T_{PQ}$ \emph{after} inflation. 
The field  acquires a VEV while its phase is identified with the QCD axion, i.e.~$\Phi= v e^{i a(t,x)/v}$, and string defects are formed (see e.g.~\cite{Vilenkin:2000jqa}). Below $T_{PQ}$, the axion evolution is:
\begin{itemize}
	\item[$\mathbf{1}$]{Most of the energy density in the strings dilutes as $\rho_{\text{strings}}\sim \mu_{s} H^{2}$, where $\mu_{s}$ is the string 
tension.~\footnote{See however~\cite{Gorghetto:2018myk, Kawasaki:2018bzv} for recent claims of small logarithmic deviations from such scaling regime.} In addition, the strings radiate axions~\cite{Kawasaki:2014sqa, Klaer:2017ond, Gorghetto:2018myk, Kawasaki:2018bzv}. 
Away from the strings, the homogeneous axion field is frozen because of Hubble friction.}

\item[$\mathbf{2}$]{At $T\lesssim O(\text{GeV})$ the QCD phase transition occurs. Non-perturbative effects generate a periodic potential for $a$ 
\begin{equation}
\label{eq:dwpotential}
	V(a, T)=\frac{m^{2}(T)v^{2}}{N_{\text{DW}}^{2}}\left[1-\cos\left(N_{\text{DW}}\frac{a}{v}\right)\right],
\end{equation}
		where $N_{\text{DW}}$ is the model dependent color anomaly, also known as \emph{DW number}. The periodicity of $	V$ is $2\pi F\equiv 2\pi v/N_{\text{DW}}$.
The  dependence of the axion mass $m(T)$ with  temperature can be parametrized as:
\begin{equation}
\label{eq:axionmass}
  m^{2}(T)=
  \begin{cases}
	  m_{0}^{2}, & \text{if}~T\lesssim T_{0},\\
	  m_{0}^{2} \left(\frac{T}{T_{0}}\right)^{-n}, & \text{if}~T\gtrsim T_{0},
  \end{cases}
\end{equation}
where $n\approx 7, T_{0}\simeq 100~\text{MeV}$ are numerical parameters which we take from~\cite{Kawasaki:2014sqa} (see~\cite{Wantz:2009it} for the original computation, \cite{Borsanyi:2016ksw}~for lattice results and the Appendix). Here, $m_{0}\simeq 0.01~\Lambda_{\text{QCD}}^2/F$ is the zero-temperature axion mass, with $\Lambda_{\text{QCD}}\simeq 400~\text{MeV}$.

The potential in \eqref{eq:dwpotential} leads to the existence of DWs, with tension $\sigma(T)~\simeq 8 m(T) F^{2}$.
These become relevant at the temperature $T_{1}\sim~\text{GeV}$ defined by $3 H(T_{1})=m(T_{1})$ (see also the Appendix).

For topological reasons, each string gets attached to $N_{\text{DW}}$ DWs at $T_{1}$. Thus, a string-wall network is formed, which also contains closed structures. At the same time, the homogeneous component of the axion field starts to oscillate and generates CDM (misalignment mechanism).}

\item[$\mathbf{3}$]{Below $T_{1}$, the energy density of the network is quickly dominated by horizon-size DWs. The subsequent evolution crucially depends on the DW number~(see e.g.~\cite{Vilenkin:2000jqa}). If $N_{\text{DW}}=1$ the network is unstable and rapidly decays. If instead $N_{\text{DW}}>1$, the network is stable because strings are pulled in different directions by the DWs. One thus faces the DW problem~\cite{Zeldovich:1974uw}. To avoid this catastrophe, a bias term can be added to the axion potential~\cite{Sikivie:1982qv},~\cite{ Hiramatsu:2010yn, Kawasaki:2014sqa} of the form:~\footnote{For the time being, we consider a bias term which switches on at $T_{0}$ and remains constant thereafter. We discuss more about this point later on.}
\begin{equation}
\label{eq:bias}
V_{B}(a)=\mathcal{A}_{B}^4\left[1-\cos\left(\frac{a}{v}+\delta\right)\right].
\end{equation}
Notice that the periodicity of \eqref{eq:bias} is different from \eqref{eq:dwpotential}; there is only one global minimum per period $2\pi v$. Furthermore, the phase $\delta$ represents a generic offset between the bias term and the QCD potential. The addition of \eqref{eq:bias} to \eqref{eq:dwpotential} leads to an energy difference between the false and true minima, $\Delta\rho\simeq \mathcal{A}_{B}^{4}$. This generates pressure which competes against the wall tension and renders the network unstable~\cite{Sikivie:1982qv}. Balance between the two competing effects is obtained when $\sigma\simeq \mathcal{A}_{B}^{4} H^{-1}$, which is confirmed by detailed numerical analysis of the network evolution in the presence of a bias term~\cite{Kawasaki:2014sqa}. Most of the network disappears at a temperature:
\begin{equation}
\label{eq:T2}
T_{2}=\epsilon \left(\frac{M_{p}\Delta \rho}{\sigma}\right)^{1/2}\left(\frac{90}{\pi^2 g_{\text{eff}}(T_{2})}\right)^{1/4},
\end{equation}
where $\epsilon\sim O(0.1-1)$ is a parameter which increases with $N_{\text{DW}}$ and has been numerically determined in~\cite{Kawasaki:2014sqa}. Depending on their initial size, most of the closed DWs in the network will collapse at different temperatures between $T_{1}$ and $T_{2}$. In the process, axions are radiated in such a way that the total axion DM abundance today is given by $\Omega_{a}=\Omega_{\text{mis}}+\Omega_{\text{strings}}+\Omega_{\text{network}}$. The axion DM abundance has been numerically studied in~\cite{Kawasaki:2014sqa} and we review the dependence of $\Omega_{a}$ on $F$ and $T_{2}$ in the Appendix.}
 
\end{itemize}
There are two crucial points to take from the discussion above: 1) for $N_{\text{DW}}>1$ there can be a significant separation between $T_{1}$, the temperature of network formation, and $T_{2}$, the temperature at which its annihilation is efficient, 
and 2) since \eqref{eq:bias} lifts the degeneracy of the $N_{\text{DW}}$ vacua, closed structures surrounding regions with energy $\mathcal{A}_{B}^{4}$ can exist in the network. Therefore, from now on we assume $N_{\text{DW}}>1$.

\begin{figure}
    \centering
    \includegraphics[width=0.45\textwidth]{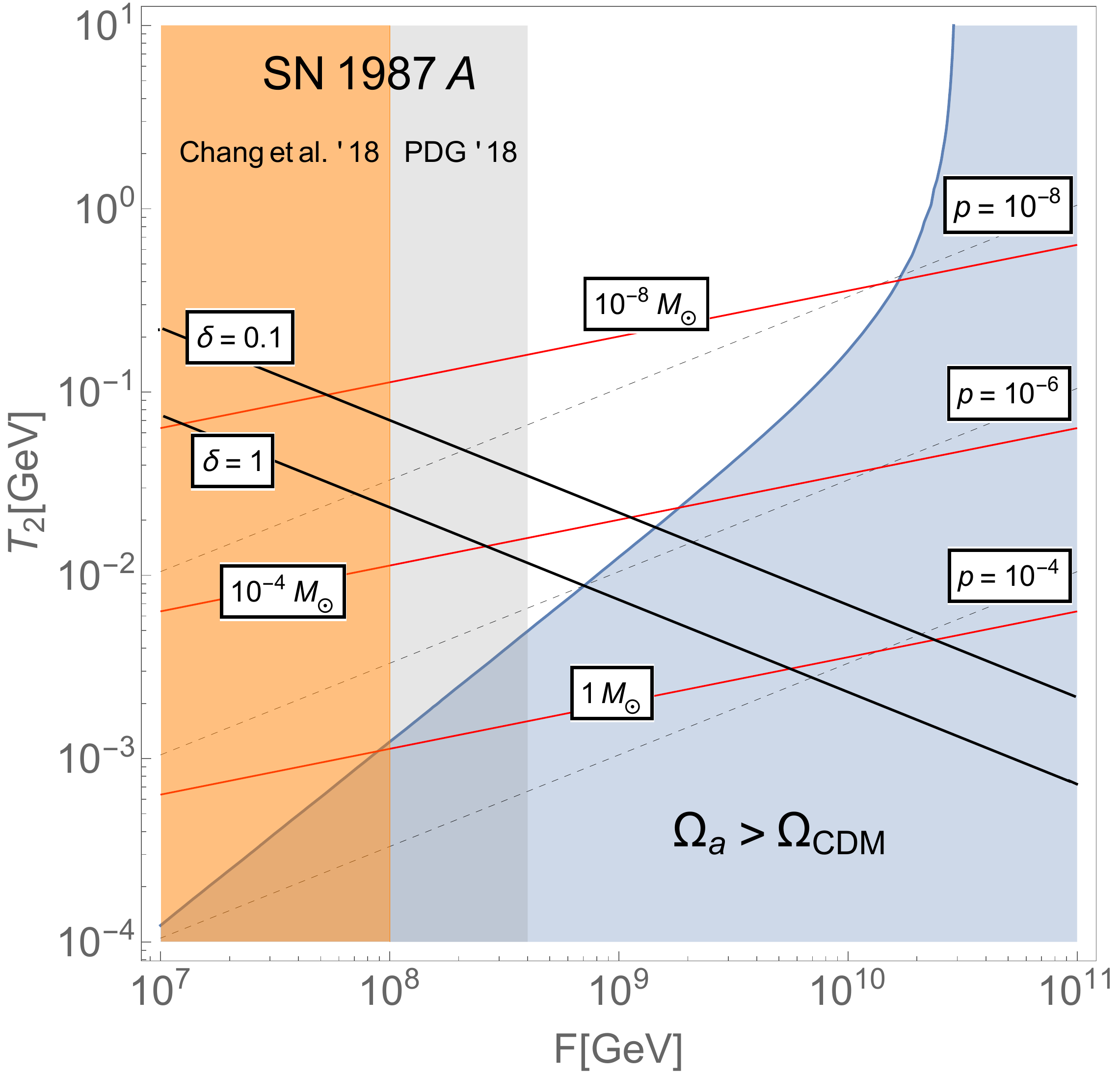}
	\caption{Constraints on $F$ and $T_{2}$ from DM overproduction (blue shaded region), and from supernovae cooling (orange shaded). The figure of merit (dashed lines) and the DW masses (red lines) are also shown. No tuning of the offset phase is required below the line $\delta=0.1$.}
    \label{fig:pspace}
\end{figure}

In Fig.~\ref{fig:pspace} we plot the constraints on the $F$-$T_{2}$ plane for $N_{\text{DW}}=2$. The blue-shaded region is excluded because of DM overproduction. In gray and orange we show the region excluded using SN-cooling arguments according to the standard analysis~\cite{Tanabashi:2018pdg} and to a more conservative recent estimate~\cite{Chang:2018rso}, respectively. The thick black lines signal the largest allowed value of the offset phase $\delta$ in \eqref{eq:bias} that does not spoil the axion solution to the strong CP problem. We thus conclude that a viable region of parameter space exists, around $T_{2}\simeq 5~\text{MeV}$ and corresponding to $\mathcal{A}_{B}\simeq 10^{-3}\Lambda_{\text{QCD}}$, where no tuning of $\delta$ is required. This is in contrast with the conclusion reached in~\cite{Kawasaki:2014sqa}, where an overconservative bound on $\theta_{\text{QCD}}$ was assumed. The untuned region of parameter space is slightly reduced as $N_{\text{DW}}$ increases. Nonetheless, even for $N_{\text{DW}}=6$ only a mild tuning $\delta\sim 0.1$ is required. 

In Fig.~\ref{fig:pspace} we also show the relevant would-be BH masses and the figure of merit for closed DWs which collapse at $T_{\star}\simeq T_{2}$. In the most interesting region of parameter space, we find $p\sim 10^{-6}$, five orders of magnitude larger than for $T_{\star}\sim T_{1}$. This shows the advantage of considering $N_{\text{DW}}>1$. Nevertheless, $p$ remains quite small and at this point it is unclear whether this leads to a significant fraction of PBHs.

\paragraph{\textbf{PBHs from late collapses.}}

Crucially, for closed DWs collapsing at $T_{\star}<T_{2}$, $p$ increases as $T_{\star}^{-4}$, because the vacuum energy contribution dominates over the wall tension, as dictated by~\eqref{eq:merit2}. 

\begin{figure}
    \centering
    \includegraphics[width=0.45\textwidth]{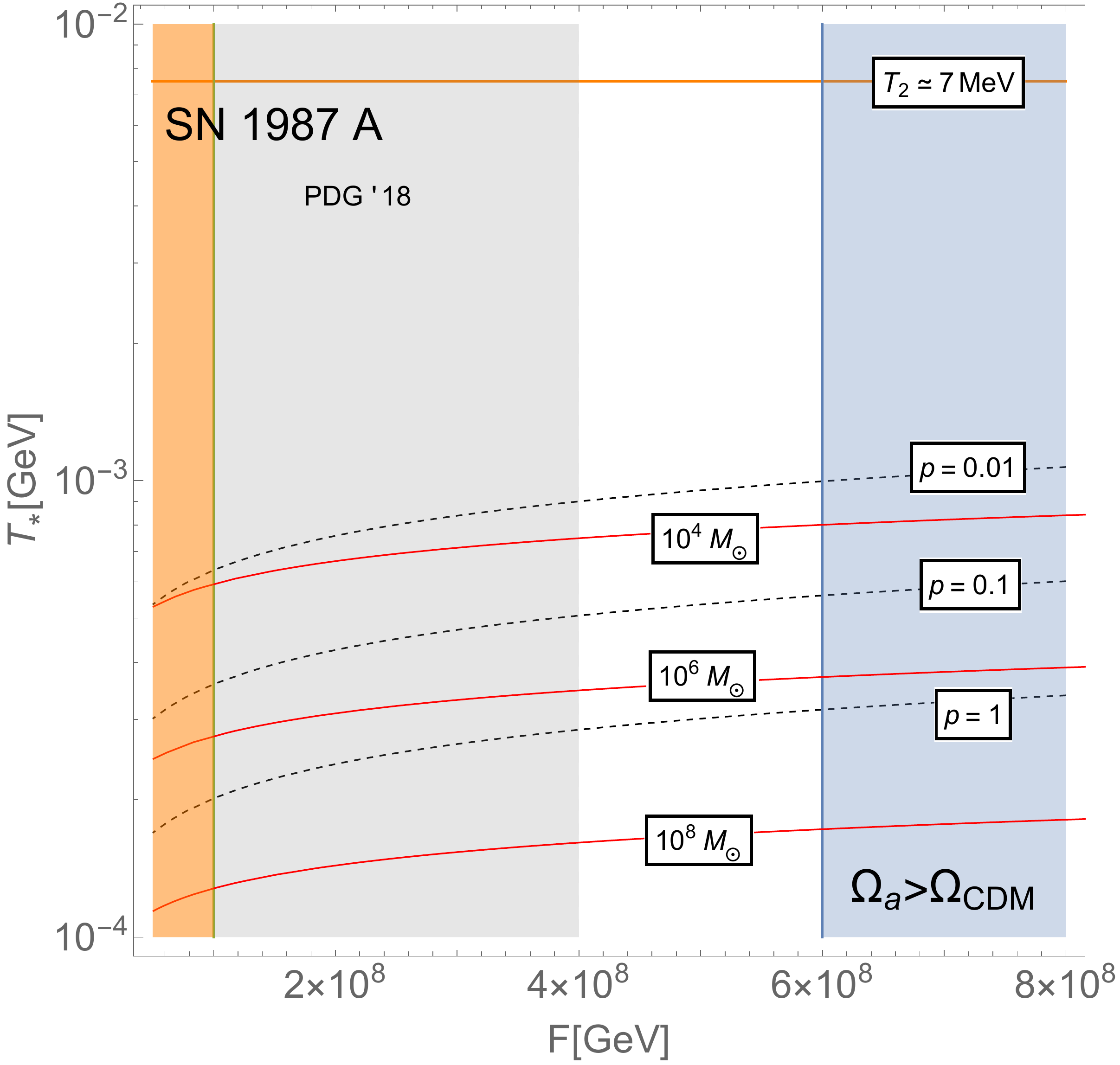}
    \caption{Constraints on $F$ for $T_{2}\simeq 7~\text{MeV}$. The figure of merit (dashed lines) and the masses of the DWs collapsing at $T_{\star}<T_{2}$ (red lines) are also shown.}
    \label{fig:late}
\end{figure}

The region around $F\lesssim 10^{9}$ GeV and $T_{2}\simeq 7$ MeV in Fig.~\ref{fig:pspace} leads to the best case scenario for PBH formation. 
In Fig.~\ref{fig:late} we plot the figure of merit and PBH masses for DWs collapsing at $T_{\star}< T_{2}$. 

Fig.~\ref{fig:late} shows that DWs collapsing roughly when $T_{\star}\sim 0.1~T_{2}$ quite likely form PBHs. These structures only have to contract by one order of magnitude before entering their Schwarzschild radius. Energy losses via radiation of axions as well as the growth of asphericities can be neglected for such short contractions. Indeed, the radiation of energy from a closed spherical Sine-Gordon DW was studied in~\cite{Widrow:1989vj} and shown to become relevant only once the wall has contracted to a size $R\sim R_{\star}^{2/3}m^{-1/3 }\ll 0.1~H_{\star}^{-1}$. Similarly, in the thin wall approximation asphericities do not spoil the PBH formation for large $p$~\cite{Widrow:1989fe}. Furthermore, we have numerically simulated the collapse of Sine-Gordon non-spherical DWs and checked that they can indeed contract down to $r_{min}\lesssim 0.1~R_{\star}$~\cite{Ferrer:2018}.~\footnote{We have neglected angular momentum in the numerical simulation. We expect that for large $p$ the inclusion of the latter should not significantly alter our picture.} The resulting PBHs would have masses $M_{\star}\sim 10^{4}-10^{7} M_{\odot}$. 

Let us estimate the fraction $f \equiv \Omega_{PBH}/\Omega_{CDM}$. After $T_{2}$, the energy density of the network is dominated by the bias contribution. However, at any given $T_{\star}<T_{2}$ only  a small fraction $P_{\text{nw}}$ of the original network survives. Therefore,
\begin{equation}
\rho_{\text{nw}}(T_{\star})\sim P_{\text{nw}}(T_{\star})\Delta\rho.
\end{equation}
Assuming that the PBH formation occurs mostly at a single temperature $T_{\star}$, the actual fraction $f$ is then given by:
\begin{equation}
\label{eq:fraction}
f\sim p^{N} \times \frac{\rho_{\text{nw}}(T_{\star})}{\rho_{\text{CDM}}(T_{\star})},
\end{equation}
where $N\gtrsim 1$ takes into account the effects of asphericities and angular momentum. One should keep in mind that $f$ might be further suppressed by the probability of finding closed structures in the network. In \eqref{eq:fraction}, $\rho_{\text{CDM}}(T_{\star})\sim\rho_{\text{CDM}}(T_{2}) (T_{\star}/T_{2})^{3}$ is the energy density of CDM at $T_{\star}$. In the most interesting region of parameter space in Fig.~\ref{fig:pspace}, $\rho_{\text{CDM}}(T_{2})$ is dominated by the contribution from axions radiated by the network. Hence, $\rho_{\text{CDM}}(T_{2})\approx \rho_{\text{nw}}(T_{2})\sim \Delta\rho$. Putting everything together, we find:
\begin{equation}
\label{eq:fraction1}
f\sim p^{N} P_{\text{nw}}(T_{\star}) \left(\frac{T_{2}}{T_{\star}}\right)^{3}.
\end{equation}
To estimate the actual value of $f$ requires knowledge of $P_{\text{nw}}$. In this respect, the simulations of~\cite{Kawasaki:2014sqa} show that, for $N_{\text{DW}}=2$, at $T_{2}$ defined by \eqref{eq:T2} with $\epsilon\simeq 0.5$ only $10\%$ of the original network survives. The percentage is further reduced to $1\%$ at $T_{2}/\sqrt{2}$ approximately. We do not know the subsequent evolution of the network. Nevertheless, let us assume for simplicity that the network decay follows a power law beyond $T_{2}$:~\footnote{In support of this choice, numerical simulations exist~\cite{Vilenkin:2000jqa}, which show that at formation the number density of closed string loops of radius $R$ scales as $R^{-3}$. Thus the density of horizon-size loops scales as $H^{-3}\sim T^{-6}$, i.e. as a power law in $T$.} 
\begin{equation}
\label{eq:extrapolation}
P_{\text{nw}}(T_{\star})\sim \left(\frac{T_{2}}{T_{\star}}\right)^{-\alpha}.
\end{equation}
Fitting~\eqref{eq:extrapolation} to the aforementioned results of \cite{Kawasaki:2014sqa} gives $\alpha\approx 7$. The final fraction $f$ then does not depend significantly on $N$. As long as $N>1$, the right hand side of~\eqref{eq:fraction1} increases as $T_{\star}$ decreases, until $p$ is saturated to one. This occurs at $T_{\star}\sim T_{2}/30$ (see Fig.~\ref{fig:late}), which gives $f\sim 10^{-6}$ and $M_{\star}\sim 10^{6} M_{\odot}$. Below this temperature, the fraction falls rapidly.

However, this result is sensitive to the precise numerical scaling of $P_{\text{nw}}$ after $T_{2}$. In this regard, it is interesting to notice that numerical simulations hint at slight deviations from the \emph{scaling regime}~\cite{Kawasaki:2014sqa}. The decay of the network can then be slower, resulting in smaller $\alpha$ and larger $f$.

Observations require $f\lesssim 10^{-5}$ for PBHs with $M\sim 10^{6} M_{\odot}$~\cite{Carr:2016drx} (see also~\cite{Carr:2018rid, Poulin:2017bwe}). This constraint is easily satisfied in our scenario. 

Nevertheless, to confidently estimate the actual fraction requires additional numerical studies, which we leave for future work. Let us remark that if a larger $f$ can be obtained with our mechanism, then (non)observations of PBHs may actually give additional constraints on axion models with $N_{DW}>1$.

\paragraph{\textbf{Origin of the bias term.}}

A minimalistic option to generate \eqref{eq:bias} is from gravity. As pointed out in \cite{Rai:1992xw} (see also \cite{Ringwald:2015dsf}), Planck-suppressed effective operators  could lead to \eqref{eq:bias}. These operators were originally investigated in~\cite{Kamionkowski:1992mf, Holman:1992us, Barr:1992qq}, which showed that they must come with very small coefficients in order not to spoil the axion solution to the strong CP problem. However, the lore is that gravity affects the PQ symmetry only at the non-perturbative level~\cite{Kallosh:1995hi} (see also~\cite{Dvali:2005an, Alonso:2017avz}). As a result, the size of the induced potential from gravity can be estimated $\sim\mathcal{A}_{B}\sim M_{p} e^{-\# M_{p}/F}$, which is certainly small albeit too small to give a viable cosmology.

Here, we propose an alternative possibility to generate the bias term. Consider a dark gauge sector, which also breaks the $U(1)_{\text{PQ}}$ via anomalies and has DW number $N_{\text{DW}}^{\text{dark}}=1$. 
The specific matter spectrum and couplings of this hidden sector are not crucial to our discussion, even though cosmological and collider constraints should be checked in any concrete realization. Such a dark sector would then precisely generate a contribution to the axion potential of the form \eqref{eq:bias}, with $\mathcal{A}_{B}$ related to the scale of dark gluon condensation, and $\delta$ containing the dark sector $\theta$-term.

Interestingly, this naturally allows for the scale $\mathcal{A}_{B}$ to have a T-dependence analogous to the QCD axion potential
\begin{equation}
\mathcal{A}_{B}(T)^{4}= m^{2}_{B}(T)v^{2}=m^{2}_{B}(T)N_{DW}^{2}F^{2},
\end{equation}
with~(see also the Appendix)
\begin{equation}
m_{B}^{2}(T)=d_{T}\frac{\Lambda_{B}^{4}}{F^{2}}\left(\frac{T}{\Lambda_{B}}\right)^{-n'}, \quad \text{if}~T\gtrsim T_{0, B}.
\end{equation}
The natural expectation is that $m_{B}$ will increase as temperature decreases until $T_{0, B}\sim \Lambda_{B}$, and remain constant afterwards. Here, $\Lambda_{B}$ is the dark confinement scale and $d_{T}, n'$ depend on the dark spectrum.

These parameters have an important impact on PBH formation. For instance, for $T_2\sim 5~\text{MeV}$ and $d_{T}, n' \sim 1$ the bias term has not yet reached its asymptotic value at $T_{2}$. Therefore, $p$ and $M_{\star}$ now scale as $T_{\star}^{-4-n'}$ and $T_{\star}^{-6-n'}$ respectively from $T_{2}$ to $T_{\star}\sim T_{0, B}$. A large figure of merit can then be attained in less than one order of magnitude in $T$, and lighter PBHs may be generated (down to $\sim 10^{4} M_{\odot}$).~\footnote{The lines of constant $\delta$ in Fig.~\ref{fig:pspace} get modified, but viable regions with $\delta\gtrsim 0.1$ persist.}
Alternatively, if $d_{T}\ll 1$ and/or $n'\gtrsim 6$, $\Lambda_{B}$ roughly coincides with $T_{2}$ and we recover the previous case.  In general, the dark sector confinement scale should be $100~\text{keV}\lesssim\Lambda_{B}\lesssim T_{2}$ in order for the mechanism presented here to generate an interesting and viable fraction of PBHs, as discussed in the previous section. We leave a more detailed investigation of the dark sector for future work.

\paragraph{\textbf{Conclusions.}}

We have discussed a new mechanism to generate PBHs in the context of QCD axion models. It proceeds by the late collapses of closed DWs in a long-lived string-DW network, which arises in QCD axion realizations with $N_{DW}>1$ and PQ symmetry broken after inflation.

Lacking accurate knowledge of the network evolution and collapse, we cannot give precise predictions for the fraction and masses of the PBHs. However, under reasonable assumptions, depending on the temperature behavior of the bias term, PBHs with masses in the range $M\sim 10^{4}-10^{7} M_{\odot}$ and representative fraction $f\gtrsim 10^{-6}$ can be created. Interestingly, such heavy PBHs can play an important role as seeds for the formation of cosmological structure, alleviating several problems of the CDM scenario on sub-galactic scales, and providing an avenue to explain the origin of the super-massive BHs~\cite{Clesse:2015wea,Carr:2018rid}.

Our proposal appears to prefer small values of the axion decay constant, $F\lesssim 10^{9}~\text{GeV}$, corresponding to axion masses in the meV range. These values are close to the lower bounds from the cooling of supernovae~\cite{Tanabashi:2018pdg, Chang:2018rso, Hamaguchi:2018oqw}, which are however subject to astrophysical uncertainties and are not universal (see e.g.~\cite{DiLuzio:2017ogq}). On the other hand, small values of $F$ might be observationally interesting. In this respect, it is intriguing that several stellar systems show a mild preference for 
a non-standard cooling mechanism, which can be interpreted in terms of a DFSZ QCD axion~\cite{Giannotti:2015kwo,Giannotti:2017hny}. In addition, several experiments will be probing this region of QCD axion masses in the near future. In particular: IAXO~\cite{Armengaud:2014gea}, TASTE~\cite{Anastassopoulos:2017kag}, ALPS  II~\cite{Bahre:2013ywa} and ARIADNE~\cite{Arvanitaki:2014dfa}.

Our mechanism might be probed at gravitational wave observatories via the detection of gravitational radiation from: SMBH binaries at LISA~\cite{Klein:2015hvg}, the annihilation of the string-wall network~\cite{Saikawa:2017hiv} at aLIGO (O5)~\cite{TheLIGOScientific:2014jea}, LISA, ET~\cite{Punturo:2010zz} and SKA~\cite{Janssen:2014dka}.

Finally, let us mention that considering very light generic Axion-Like-Particles, the network collapse could be delayed to $T_{\star}\lesssim 100~\text{keV}$ raising the figure of merit. The resulting extremely heavy PBHs, however, are strongly constrained.

\begin{acknowledgments}

\paragraph{\textbf{Acknowledgments.}} We thank J.J. Blanco-Pillado, J. Garriga, J. Redondo, K. Saikawa, G. Servant and T. Vachaspati for useful discussions. We acknowledge support by the Spanish Ministry MEC under grant FPA2014-55613-P and the Severo Ochoa excellence program of MINECO (grant SO-2012- 0234, SEV-2016- 0588), as well as by the Generalitat de Catalunya under grant 2014-SGR-1450.
F. F. was also supported in part by the U.S. Department of Energy, Office of High Energy Physics, under Award No. DE-FG02-91ER40628 and \text{DE-SC0017987}.

\end{acknowledgments}

\begin{appendices}

\section{Appendix: QCD axion dark matter}

The aim of this appendix is to review the relevant formulae for the total axion dark matter abundance. The material presented here can be partially found in~\cite{Kawasaki:2014sqa}~(and refs. therein), together with more detailed explanations.

The total axion dark matter abundance is given by
\begin{equation}
\label{eq:totdmap}
\Omega_{a}=\Omega_{\text{mis}}+\Omega_{\text{strings}}+\Omega_{\text{nw}},
\end{equation}
where the three terms on the right hand side of \eqref{eq:totdmap} represent respectively the contribution from: the misalignment mechanism, the radiation from axionic strings and the radiation from the string-wall network.

Let us first provide formulae for the axion mass. Following~\cite{Kawasaki:2014sqa}, we have
\begin{equation}
\label{eq:axionmass}
  m^{2}(T)=
  \begin{cases}
	  c_{0}\frac{\Lambda_{\text{QCD}}^{4}}{F^{2}}, & \text{if}~T\lesssim T_{0},\\
	  c_{T}\frac{\Lambda_{\text{QCD}}^{4}}{F^{2}} \left(\frac{T}{\Lambda_{\text{QCD}}}\right)^{-n}, & \text{if}~T\gtrsim T_{0}.
  \end{cases}
\end{equation}
The parameters $c_{0}, c_{T}$ and $n$ can be determined using the Dilute Instanton Gas Approximation (DIGA)~(see e.g.~\cite{Coleman:1978ae}), valid at high temperatures. We take $c_{0}\approx 10^{-3}, c_{T}\approx 10^{-7}$ and $n\approx 7$ following~\cite{Wantz:2009it}~(see also lattice QCD results which agree~\cite{Borsanyi:2016ksw, Borsanyi:2015cka} or deviate~\cite{Bonati:2016imp} from these values). From \eqref{eq:axionmass} one finds $T_{0}\simeq~ 100~\text{MeV}$.

Let us now move to the relic abundance. Firstly, let us focus on the contribution from the misalignment mechanism $\Omega_{\text{mis}}$. The QCD axion starts to oscillate at the temperature $T_{1}$ given by $3H(T_{1})=m(T_{1})$. By means of \eqref{eq:axionmass}, we find
\begin{equation}
\label{eq:T1ap}
T_{1}\simeq A_{n} \left(\frac{g_{\text{\text{eff}}}}{80}\right)^{-\frac{1}{4+n}}\left(\frac{F}{10^{9}~\text{GeV}}\right)^{-\frac{2}{4+n}}\Lambda_{QCD},
\end{equation}
where  $A_{n}=(7.5\cdot 10^{16} c_{T})^{\frac{1}{4+n}}$. For $c_{T}\approx 10^{-7}$ and $n\approx 7$, this gives $T_1\approx 3$~GeV for $F\simeq 10^{9}\text{GeV}$. The relic abundance is given by
\begin{align}
\nonumber \Omega_{\text{mis}}h^{2} &\simeq B_{n}\sqrt{c_{0}} c_{T}^{-\frac{1}{4+n}}\left(\frac{F}{10^{9}~\text{GeV}}\right)^{\frac{6+n}{4+n}}\\
\label{eq:misab}
&\times \left(\frac{g_{\text{eff}}(T_{1})}{80}\right)^{-\frac{6+n}{2(4+n)}}\left(\frac{\Lambda_{QCD}}{400~\text{MeV}}\right)
\end{align}
$B_{n}\simeq 0.8 \cdot 10^{9} \times (2.2\cdot 10^{10})^{-\frac{6+n}{4+n}}$. Using the DIGA values for these parameters, the misalignment contribution saturates the observed dark matter abundance for $F\simeq 10^{11}~\text{GeV}$.

Let us now move on to $\Omega_{\text{strings}}$. At the moment, there is some controversy in the literature regarding the magnitude of this contribution (see~\cite{Klaer:2017ond, Gorghetto:2018myk, Kawasaki:2018bzv} for recent estimates with a different take on previous calculations). Without entering into details, we focus on the parametric dependence of  $\Omega_{\text{strings}}$ on $F$
\begin{align}
\nonumber \Omega_{\text{strings}}h^{2}&\simeq C_{n} \left(\frac{F}{10^9~\text{GeV}}\right)^{\frac{6+n}{4+n}}\\
\label{eq:stringsab}
\times &\left(\frac{g_{\text{eff}}(T_{1})}{80}\right)^{-\frac{2+n}{2(4+n)}}\left(\frac{\Lambda_{QCD}}{400~\text{MeV}}\right),
\end{align}
where $C_{n}$ is a numerical prefactor. In order to produce Fig.~\ref{fig:pspace}, we have used the formulae provided in~\cite{Kawasaki:2014sqa}, where $C_{n}\sim 10^{-3}$. In this case the contribution from strings is generically larger or comparable to the contribution from the misalignment angle; and the sum of the two contributions saturates the observed dark matter abundance for $F\simeq 2\times 10^{10}$ GeV. The precise behavior of $\Omega_{\text{strings}}$ is not crucial to our proposal, since we are especially interested in the region of small $F$, where $\Omega_{\text{mis}}$ and $\Omega_{\text{strings}}$ represent a subdominant contribution to the total axion abundance.

Let us finally discuss the contribution from the string-wall network, which is especially important for our proposal. The crucial difference with respect to the abundances from the misalignment mechanism and from strings is that the network radiates axions at $T_{2}<T_{1}$. Thus, their abundance is less diluted and for small values of $F$ it dominates over the other contributions. Assuming a so-called exact \emph{scaling regime} for the evolution of the network, i.e.~$\rho_{\text{nw}}\sim \sigma H$, we have
\begin{align}
\label{eq:omegadw}
\nonumber \Omega_{\text{nw}}h^{2}& \simeq 0.14\times \left(\frac{F}{10^{9}~\text{GeV}}\right)\left(\frac{\Lambda_{QCD}}{400~\text{MeV}}\right)^{2}\\ & \times\left(\frac{g_{\text{eff}}(T_{2})}{10.75}\right)^{-1/4}\left(\frac{10~\text{MeV}}{T_{2}}\right).
\end{align}
Our expression for $\Omega_{\text{nw}}$ differs from the one presented in~\cite{Kawasaki:2014sqa} in that we keep the dependence on $T_{2}$, rather than trading it for $\mathcal{A}_{B}$ according to~\eqref{eq:T2}. Furthermore, there are numerical prefactors in \eqref{eq:omegadw} which we have fixed according to the results of~\cite{Kawasaki:2014sqa}. 
The sum of \eqref{eq:misab},~\eqref{eq:stringsab} and \eqref{eq:omegadw} generates the solid blue curve in Fig.~\ref{fig:pspace}.

Let us now discuss the solid lines of constant $\delta$ in Fig.~\ref{fig:pspace}. The addition of the bias term~\eqref{eq:bias} misaligns the axion from the CP conserving minimum determined by the QCD potential~\eqref{eq:dwpotential}. In particular, the QCD angle $\theta\equiv a/F$ at the minimum is approximately given by:
\begin{equation}
\label{eq:thetamin}
\theta_{min}\simeq \frac{\mathcal{A}_{B}^4 N_{DW} \sin\delta}{m^{2}N_{DW}^2 F^{2}+\mathcal{A}_{B}^4\cos\delta}.
\end{equation}
Inverting~\eqref{eq:T2} and using $\Delta\rho\simeq \mathcal{A}_{B}^4\left[1-\cos\left(2\pi/N_{DW}\right)\right]$, we find:
\begin{equation}
\mathcal{A}_{B}^{4}\simeq \sqrt{\frac{\pi^2 g_{\text{\text{eff}}}(T_{2})}{90}}\frac{T_{2}^2}{\epsilon\left[1-\cos\left(2\pi/N_{DW}\right)\right]}\frac{\sigma}{M_{p}}.
\end{equation}
In order to preserve the solution to the strong CP problem, we require $\theta_{min}\lesssim 10^{-10}$~\cite{Tanabashi:2018pdg}. At constant $\delta$, \eqref{eq:thetamin} corresponds to a line in the logarithmic $F-T_{2}$ plane, as shown in Fig.~\ref{fig:pspace}. Notice that the position of these lines depends on $N_{DW}$: in particular, the region of phenomenologically viable parameter space where $\delta\sim 1$ shrinks as we increase $N_{DW}$. Nevertheless, even for $N_{DW}=6$ there is an allowed region of parameter space where only $\delta\sim 0.1$ is required.

\end{appendices}


\begin{thebibliography}{66}
\expandafter\ifx\csname natexlab\endcsname\relax\def\natexlab#1{#1}\fi
\expandafter\ifx\csname bibnamefont\endcsname\relax
  \def\bibnamefont#1{#1}\fi
\expandafter\ifx\csname bibfnamefont\endcsname\relax
  \def\bibfnamefont#1{#1}\fi
\expandafter\ifx\csname citenamefont\endcsname\relax
  \def\citenamefont#1{#1}\fi
\expandafter\ifx\csname url\endcsname\relax
  \def\url#1{\texttt{#1}}\fi
\expandafter\ifx\csname urlprefix\endcsname\relax\def\urlprefix{URL }\fi
\providecommand{\bibinfo}[2]{#2}
\providecommand{\eprint}[2][]{\url{#2}}

\bibitem[{\citenamefont{Abbott et~al.}(2016)}]{Abbott:2016blz}
\bibinfo{author}{\bibfnamefont{B.~P.} \bibnamefont{Abbott}}
  \bibnamefont{et~al.} (\bibinfo{collaboration}{Virgo, LIGO Scientific}),
  \bibinfo{journal}{Phys. Rev. Lett.} \textbf{\bibinfo{volume}{116}},
  \bibinfo{pages}{061102} (\bibinfo{year}{2016}), \eprint{1602.03837}.

\bibitem[{\citenamefont{Hawking}(1971)}]{Hawking:1971ei}
\bibinfo{author}{\bibfnamefont{S.}~\bibnamefont{Hawking}},
  \bibinfo{journal}{Mon. Not. Roy. Astron. Soc.}
  \textbf{\bibinfo{volume}{152}}, \bibinfo{pages}{75} (\bibinfo{year}{1971}).

\bibitem[{\citenamefont{Carr and Hawking}(1974)}]{Carr:1974nx}
\bibinfo{author}{\bibfnamefont{B.~J.} \bibnamefont{Carr}} \bibnamefont{and}
  \bibinfo{author}{\bibfnamefont{S.~W.} \bibnamefont{Hawking}},
  \bibinfo{journal}{Mon. Not. Roy. Astron. Soc.}
  \textbf{\bibinfo{volume}{168}}, \bibinfo{pages}{399} (\bibinfo{year}{1974}).

\bibitem[{\citenamefont{Bird et~al.}(2016)\citenamefont{Bird, Cholis, Muñoz,
  Ali-Haïmoud, Kamionkowski, Kovetz, Raccanelli, and Riess}}]{Bird:2016dcv}
\bibinfo{author}{\bibfnamefont{S.}~\bibnamefont{Bird}},
  \bibinfo{author}{\bibfnamefont{I.}~\bibnamefont{Cholis}},
  \bibinfo{author}{\bibfnamefont{J.~B.} \bibnamefont{Muñoz}},
  \bibinfo{author}{\bibfnamefont{Y.}~\bibnamefont{Ali-Haïmoud}},
  \bibinfo{author}{\bibfnamefont{M.}~\bibnamefont{Kamionkowski}},
  \bibinfo{author}{\bibfnamefont{E.~D.} \bibnamefont{Kovetz}},
  \bibinfo{author}{\bibfnamefont{A.}~\bibnamefont{Raccanelli}},
  \bibnamefont{and} \bibinfo{author}{\bibfnamefont{A.~G.} \bibnamefont{Riess}},
  \bibinfo{journal}{Phys. Rev. Lett.} \textbf{\bibinfo{volume}{116}},
  \bibinfo{pages}{201301} (\bibinfo{year}{2016}), \eprint{1603.00464}.

\bibitem[{\citenamefont{Clesse and García-Bellido}(2017)}]{Clesse:2016vqa}
\bibinfo{author}{\bibfnamefont{S.}~\bibnamefont{Clesse}} \bibnamefont{and}
  \bibinfo{author}{\bibfnamefont{J.}~\bibnamefont{García-Bellido}},
  \bibinfo{journal}{Phys. Dark Univ.} \textbf{\bibinfo{volume}{15}},
  \bibinfo{pages}{142} (\bibinfo{year}{2017}), \eprint{1603.05234}.

\bibitem[{\citenamefont{Sasaki et~al.}(2016)\citenamefont{Sasaki, Suyama,
  Tanaka, and Yokoyama}}]{Sasaki:2016jop}
\bibinfo{author}{\bibfnamefont{M.}~\bibnamefont{Sasaki}},
  \bibinfo{author}{\bibfnamefont{T.}~\bibnamefont{Suyama}},
  \bibinfo{author}{\bibfnamefont{T.}~\bibnamefont{Tanaka}}, \bibnamefont{and}
  \bibinfo{author}{\bibfnamefont{S.}~\bibnamefont{Yokoyama}},
  \bibinfo{journal}{Phys. Rev. Lett.} \textbf{\bibinfo{volume}{117}},
  \bibinfo{pages}{061101} (\bibinfo{year}{2016}), \eprint{1603.08338}.

\bibitem[{\citenamefont{Kashlinsky}(2016)}]{Kashlinsky:2016sdv}
\bibinfo{author}{\bibfnamefont{A.}~\bibnamefont{Kashlinsky}},
  \bibinfo{journal}{Astrophys. J.} \textbf{\bibinfo{volume}{823}},
  \bibinfo{pages}{L25} (\bibinfo{year}{2016}), \eprint{1605.04023}.

\bibitem[{\citenamefont{Carr et~al.}(2016)\citenamefont{Carr, Kuhnel, and
  Sandstad}}]{Carr:2016drx}
\bibinfo{author}{\bibfnamefont{B.}~\bibnamefont{Carr}},
  \bibinfo{author}{\bibfnamefont{F.}~\bibnamefont{Kuhnel}}, \bibnamefont{and}
  \bibinfo{author}{\bibfnamefont{M.}~\bibnamefont{Sandstad}},
  \bibinfo{journal}{Phys. Rev.} \textbf{\bibinfo{volume}{D94}},
  \bibinfo{pages}{083504} (\bibinfo{year}{2016}), \eprint{1607.06077}.

\bibitem[{\citenamefont{Clesse and García-Bellido}(2015)}]{Clesse:2015wea}
\bibinfo{author}{\bibfnamefont{S.}~\bibnamefont{Clesse}} \bibnamefont{and}
  \bibinfo{author}{\bibfnamefont{J.}~\bibnamefont{García-Bellido}},
  \bibinfo{journal}{Phys. Rev.} \textbf{\bibinfo{volume}{D92}},
  \bibinfo{pages}{023524} (\bibinfo{year}{2015}), \eprint{1501.07565}.

\bibitem[{\citenamefont{Carr and Silk}(2018)}]{Carr:2018rid}
\bibinfo{author}{\bibfnamefont{B.}~\bibnamefont{Carr}} \bibnamefont{and}
  \bibinfo{author}{\bibfnamefont{J.}~\bibnamefont{Silk}}
  (\bibinfo{year}{2018}), \eprint{1801.00672}.

\bibitem[{\citenamefont{Wu et~al.}(2015)}]{Wu:2015}
\bibinfo{author}{\bibfnamefont{X.-B.} \bibnamefont{Wu}} \bibnamefont{et~al.},
  \bibinfo{journal}{Nature} \textbf{\bibinfo{volume}{518}},
  \bibinfo{pages}{512} (\bibinfo{year}{2015}), \eprint{1502.07418}.

\bibitem[{\citenamefont{Banados et~al.}(2018)}]{Banados:2017unc}
\bibinfo{author}{\bibfnamefont{E.}~\bibnamefont{Banados}} \bibnamefont{et~al.},
  \bibinfo{journal}{Nature} \textbf{\bibinfo{volume}{553}},
  \bibinfo{pages}{473} (\bibinfo{year}{2018}), \eprint{1712.01860}.

\bibitem[{\citenamefont{Vilenkin and Shellard}(2000)}]{Vilenkin:2000jqa}
\bibinfo{author}{\bibfnamefont{A.}~\bibnamefont{Vilenkin}} \bibnamefont{and}
  \bibinfo{author}{\bibfnamefont{E.~P.~S.} \bibnamefont{Shellard}},
  \emph{\bibinfo{title}{{Cosmic Strings and Other Topological Defects}}}
  (\bibinfo{publisher}{Cambridge University Press}, \bibinfo{year}{2000}), ISBN
  \bibinfo{isbn}{9780521654760}.

\bibitem[{\citenamefont{Vilenkin}(1981)}]{Vilenkin:1981iu}
\bibinfo{author}{\bibfnamefont{A.}~\bibnamefont{Vilenkin}},
  \bibinfo{journal}{Phys. Rev. Lett.} \textbf{\bibinfo{volume}{46}},
  \bibinfo{pages}{1169} (\bibinfo{year}{1981}), \bibinfo{note}{[Erratum: Phys.
  Rev. Lett.46,1496(1981)]}.

\bibitem[{\citenamefont{Hawking}(1990)}]{Hawking:1990tx}
\bibinfo{author}{\bibfnamefont{S.~W.} \bibnamefont{Hawking}},
  \bibinfo{journal}{Phys. Lett.} \textbf{\bibinfo{volume}{B246}},
  \bibinfo{pages}{36} (\bibinfo{year}{1990}).

\bibitem[{\citenamefont{Fort and Vachaspati}(1993)}]{Fort:1993zb}
\bibinfo{author}{\bibfnamefont{J.}~\bibnamefont{Fort}} \bibnamefont{and}
  \bibinfo{author}{\bibfnamefont{T.}~\bibnamefont{Vachaspati}},
  \bibinfo{journal}{Phys. Lett.} \textbf{\bibinfo{volume}{B311}},
  \bibinfo{pages}{41} (\bibinfo{year}{1993}), \eprint{hep-th/9305081}.

\bibitem[{\citenamefont{Garriga and Sakellariadou}(1993)}]{Garriga:1993gj}
\bibinfo{author}{\bibfnamefont{J.}~\bibnamefont{Garriga}} \bibnamefont{and}
  \bibinfo{author}{\bibfnamefont{M.}~\bibnamefont{Sakellariadou}},
  \bibinfo{journal}{Phys. Rev.} \textbf{\bibinfo{volume}{D48}},
  \bibinfo{pages}{2502} (\bibinfo{year}{1993}), \eprint{hep-th/9303024}.

\bibitem[{\citenamefont{Khlopov et~al.}(2005)\citenamefont{Khlopov, Rubin, and
  Sakharov}}]{Khlopov:2004sc}
\bibinfo{author}{\bibfnamefont{M.~{\relax Yu}.} \bibnamefont{Khlopov}},
  \bibinfo{author}{\bibfnamefont{S.~G.} \bibnamefont{Rubin}}, \bibnamefont{and}
  \bibinfo{author}{\bibfnamefont{A.~S.} \bibnamefont{Sakharov}},
  \bibinfo{journal}{Astropart. Phys.} \textbf{\bibinfo{volume}{23}},
  \bibinfo{pages}{265} (\bibinfo{year}{2005}), \eprint{astro-ph/0401532}.

\bibitem[{\citenamefont{Deng et~al.}(2017)\citenamefont{Deng, Garriga, and
  Vilenkin}}]{Deng:2016vzb}
\bibinfo{author}{\bibfnamefont{H.}~\bibnamefont{Deng}},
  \bibinfo{author}{\bibfnamefont{J.}~\bibnamefont{Garriga}}, \bibnamefont{and}
  \bibinfo{author}{\bibfnamefont{A.}~\bibnamefont{Vilenkin}},
  \bibinfo{journal}{JCAP} \textbf{\bibinfo{volume}{1704}}, \bibinfo{pages}{050}
  (\bibinfo{year}{2017}), \eprint{1612.03753}.

\bibitem[{\citenamefont{Kibble et~al.}(1982)\citenamefont{Kibble, Lazarides,
  and Shafi}}]{Kibble:1982dd}
\bibinfo{author}{\bibfnamefont{T.~W.~B.} \bibnamefont{Kibble}},
  \bibinfo{author}{\bibfnamefont{G.}~\bibnamefont{Lazarides}},
  \bibnamefont{and} \bibinfo{author}{\bibfnamefont{Q.}~\bibnamefont{Shafi}},
  \bibinfo{journal}{Phys. Rev.} \textbf{\bibinfo{volume}{D26}},
  \bibinfo{pages}{435} (\bibinfo{year}{1982}).

\bibitem[{\citenamefont{Vilenkin and Everett}(1982)}]{Vilenkin:1982ks}
\bibinfo{author}{\bibfnamefont{A.}~\bibnamefont{Vilenkin}} \bibnamefont{and}
  \bibinfo{author}{\bibfnamefont{A.~E.} \bibnamefont{Everett}},
  \bibinfo{journal}{Phys. Rev. Lett.} \textbf{\bibinfo{volume}{48}},
  \bibinfo{pages}{1867} (\bibinfo{year}{1982}).

\bibitem[{\citenamefont{Vachaspati}(2017)}]{Vachaspati:2017hjw}
\bibinfo{author}{\bibfnamefont{T.}~\bibnamefont{Vachaspati}}
  (\bibinfo{year}{2017}), \eprint{1706.03868}.

\bibitem[{\citenamefont{Sakharov and Khlopov}(1994)}]{Sakharov:1994id}
\bibinfo{author}{\bibfnamefont{A.~S.} \bibnamefont{Sakharov}} \bibnamefont{and}
  \bibinfo{author}{\bibfnamefont{M.~{\relax Yu}.} \bibnamefont{Khlopov}},
  \bibinfo{journal}{Phys. Atom. Nucl.} \textbf{\bibinfo{volume}{57}},
  \bibinfo{pages}{485} (\bibinfo{year}{1994}), \bibinfo{note}{[Yad.
  Fiz.57,514(1994)]}.

\bibitem[{\citenamefont{Sakharov et~al.}(1996)\citenamefont{Sakharov, Sokoloff,
  and Khlopov}}]{Sakharov:1996xg}
\bibinfo{author}{\bibfnamefont{A.~S.} \bibnamefont{Sakharov}},
  \bibinfo{author}{\bibfnamefont{D.~D.} \bibnamefont{Sokoloff}},
  \bibnamefont{and} \bibinfo{author}{\bibfnamefont{M.~{\relax Yu}.}
  \bibnamefont{Khlopov}}, \bibinfo{journal}{Phys. Atom. Nucl.}
  \textbf{\bibinfo{volume}{59}}, \bibinfo{pages}{1005} (\bibinfo{year}{1996}),
  \bibinfo{note}\bibinfo{note}{[Yad.
  Fiz.59N6,1050(1996)]}.

\bibitem[{\citenamefont{Khlopov et~al.}(1999)\citenamefont{Khlopov, Sakharov,
  and Sokoloff}}]{Khlopov:1999tm}
\bibinfo{author}{\bibfnamefont{M.~{\relax Yu}.} \bibnamefont{Khlopov}},
  \bibinfo{author}{\bibfnamefont{A.~S.} \bibnamefont{Sakharov}},
  \bibnamefont{and} \bibinfo{author}{\bibfnamefont{D.~D.}
  \bibnamefont{Sokoloff}}, \bibinfo{journal}{Nucl. Phys. Proc. Suppl.}
  \textbf{\bibinfo{volume}{72}}, \bibinfo{pages}{105} (\bibinfo{year}{1999}).

\bibitem[{\citenamefont{Kim}(1979)}]{Kim:1979if}
\bibinfo{author}{\bibfnamefont{J.~E.} \bibnamefont{Kim}},
  \bibinfo{journal}{Phys. Rev. Lett.} \textbf{\bibinfo{volume}{43}},
  \bibinfo{pages}{103} (\bibinfo{year}{1979}).

\bibitem[{\citenamefont{Shifman et~al.}(1980)\citenamefont{Shifman, Vainshtein,
  and Zakharov}}]{Shifman:1979if}
\bibinfo{author}{\bibfnamefont{M.~A.} \bibnamefont{Shifman}},
  \bibinfo{author}{\bibfnamefont{A.~I.} \bibnamefont{Vainshtein}},
  \bibnamefont{and} \bibinfo{author}{\bibfnamefont{V.~I.}
  \bibnamefont{Zakharov}}, \bibinfo{journal}{Nucl. Phys.}
  \textbf{\bibinfo{volume}{B166}}, \bibinfo{pages}{493} (\bibinfo{year}{1980}).

\bibitem[{\citenamefont{Dine et~al.}(1981)\citenamefont{Dine, Fischler, and
  Srednicki}}]{Dine:1981rt}
\bibinfo{author}{\bibfnamefont{M.}~\bibnamefont{Dine}},
  \bibinfo{author}{\bibfnamefont{W.}~\bibnamefont{Fischler}}, \bibnamefont{and}
  \bibinfo{author}{\bibfnamefont{M.}~\bibnamefont{Srednicki}},
  \bibinfo{journal}{Phys. Lett.} \textbf{\bibinfo{volume}{104B}},
  \bibinfo{pages}{199} (\bibinfo{year}{1981}).

\bibitem[{\citenamefont{Zhitnitsky}(1980)}]{Zhitnitsky:1980tq}
\bibinfo{author}{\bibfnamefont{A.~R.} \bibnamefont{Zhitnitsky}},
  \bibinfo{journal}{Sov. J. Nucl. Phys.} \textbf{\bibinfo{volume}{31}},
  \bibinfo{pages}{260} (\bibinfo{year}{1980}), \bibinfo{note}{[Yad.
  Fiz.31,497(1980)]}.

\bibitem[{\citenamefont{Peccei and Quinn}(1977)}]{Peccei:1977hh}
\bibinfo{author}{\bibfnamefont{R.~D.} \bibnamefont{Peccei}} \bibnamefont{and}
  \bibinfo{author}{\bibfnamefont{H.~R.} \bibnamefont{Quinn}},
  \bibinfo{journal}{Phys. Rev. Lett.} \textbf{\bibinfo{volume}{38}},
  \bibinfo{pages}{1440} (\bibinfo{year}{1977}), \bibinfo{note}{[,328(1977)]}.

\bibitem[{\citenamefont{Wilczek}(1978)}]{Wilczek:1977pj}
\bibinfo{author}{\bibfnamefont{F.}~\bibnamefont{Wilczek}},
  \bibinfo{journal}{Phys. Rev. Lett.} \textbf{\bibinfo{volume}{40}},
  \bibinfo{pages}{279} (\bibinfo{year}{1978}).

\bibitem[{\citenamefont{Weinberg}(1978)}]{Weinberg:1977ma}
\bibinfo{author}{\bibfnamefont{S.}~\bibnamefont{Weinberg}},
  \bibinfo{journal}{Phys. Rev. Lett.} \textbf{\bibinfo{volume}{40}},
  \bibinfo{pages}{223} (\bibinfo{year}{1978}).
  
\bibitem[{\citenamefont{Zeldovich et~al.}(1974)\citenamefont{Zeldovich,
  Kobzarev, and Okun}}]{Zeldovich:1974uw}
\bibinfo{author}{\bibfnamefont{{\relax Ya}.~B.} \bibnamefont{Zeldovich}},
  \bibinfo{author}{\bibfnamefont{I.~{\relax Yu}.} \bibnamefont{Kobzarev}},
  \bibnamefont{and} \bibinfo{author}{\bibfnamefont{L.~B.} \bibnamefont{Okun}},
  \bibinfo{journal}{Zh. Eksp. Teor. Fiz.} \textbf{\bibinfo{volume}{67}},
  \bibinfo{pages}{3} (\bibinfo{year}{1974}), \bibinfo{note}{[Sov. Phys.
  JETP40,1(1974)]}.

\bibitem[{\citenamefont{Sikivie}(1982)}]{Sikivie:1982qv}
\bibinfo{author}{\bibfnamefont{P.}~\bibnamefont{Sikivie}},
  \bibinfo{journal}{Phys. Rev. Lett.} \textbf{\bibinfo{volume}{48}},
  \bibinfo{pages}{1156} (\bibinfo{year}{1982}).

\bibitem[{\citenamefont{Klaer and Moore}(2017)}]{Klaer:2017ond}
\bibinfo{author}{\bibfnamefont{V.~B.} \bibnamefont{Klaer}} \bibnamefont{and}
  \bibinfo{author}{\bibfnamefont{G.~D.} \bibnamefont{Moore}},
  \bibinfo{journal}{JCAP} \textbf{\bibinfo{volume}{1711}}, \bibinfo{pages}{049}
  (\bibinfo{year}{2017}), \eprint{1708.07521}.

\bibitem[{\citenamefont{Gorghetto et~al.}(2018)\citenamefont{Gorghetto, Hardy,
  and Villadoro}}]{Gorghetto:2018myk}
\bibinfo{author}{\bibfnamefont{M.}~\bibnamefont{Gorghetto}},
  \bibinfo{author}{\bibfnamefont{E.}~\bibnamefont{Hardy}}, \bibnamefont{and}
  \bibinfo{author}{\bibfnamefont{G.}~\bibnamefont{Villadoro}}
  (\bibinfo{year}{2018}), \eprint{1806.04677}.

\bibitem[{\citenamefont{Kawasaki et~al.}(2018)\citenamefont{Kawasaki,
  Sekiguchi, Yamaguchi, and Yokoyama}}]{Kawasaki:2018bzv}
\bibinfo{author}{\bibfnamefont{M.}~\bibnamefont{Kawasaki}},
  \bibinfo{author}{\bibfnamefont{T.}~\bibnamefont{Sekiguchi}},
  \bibinfo{author}{\bibfnamefont{M.}~\bibnamefont{Yamaguchi}},
  \bibnamefont{and} \bibinfo{author}{\bibfnamefont{J.}~\bibnamefont{Yokoyama}}
  (\bibinfo{year}{2018}), \eprint{1806.05566}.

\bibitem[{\citenamefont{Kawasaki et~al.}(2015)\citenamefont{Kawasaki, Saikawa,
  and Sekiguchi}}]{Kawasaki:2014sqa}
\bibinfo{author}{\bibfnamefont{M.}~\bibnamefont{Kawasaki}},
  \bibinfo{author}{\bibfnamefont{K.}~\bibnamefont{Saikawa}}, \bibnamefont{and}
  \bibinfo{author}{\bibfnamefont{T.}~\bibnamefont{Sekiguchi}},
  \bibinfo{journal}{Phys. Rev.} \textbf{\bibinfo{volume}{D91}},
  \bibinfo{pages}{065014} (\bibinfo{year}{2015}), \eprint{1412.0789}.

\bibitem[{\citenamefont{Ringwald}(2018)}]{Ringwald:2018xlf}
\bibinfo{author}{\bibfnamefont{A.}~\bibnamefont{Ringwald}}
  (\bibinfo{year}{2018}), \eprint{1805.09618}.

\bibitem[{\citenamefont{Wantz and Shellard}(2010)}]{Wantz:2009it}
\bibinfo{author}{\bibfnamefont{O.}~\bibnamefont{Wantz}} \bibnamefont{and}
  \bibinfo{author}{\bibfnamefont{E.~P.~S.} \bibnamefont{Shellard}},
  \bibinfo{journal}{Phys. Rev.} \textbf{\bibinfo{volume}{D82}},
  \bibinfo{pages}{123508} (\bibinfo{year}{2010}), \eprint{0910.1066}.

\bibitem[{\citenamefont{Borsanyi et~al.}(2016)}]{Borsanyi:2016ksw}
\bibinfo{author}{\bibfnamefont{S.}~\bibnamefont{Borsanyi}}
  \bibnamefont{et~al.}, \bibinfo{journal}{Nature}
  \textbf{\bibinfo{volume}{539}}, \bibinfo{pages}{69} (\bibinfo{year}{2016}),
  \eprint{1606.07494}.

\bibitem[{\citenamefont{Hiramatsu et~al.}(2011)\citenamefont{Hiramatsu,
  Kawasaki, and Saikawa}}]{Hiramatsu:2010yn}
\bibinfo{author}{\bibfnamefont{T.}~\bibnamefont{Hiramatsu}},
  \bibinfo{author}{\bibfnamefont{M.}~\bibnamefont{Kawasaki}}, \bibnamefont{and}
  \bibinfo{author}{\bibfnamefont{K.}~\bibnamefont{Saikawa}},
  \bibinfo{journal}{JCAP} \textbf{\bibinfo{volume}{1108}}, \bibinfo{pages}{030}
  (\bibinfo{year}{2011}), \eprint{1012.4558}.

\bibitem[{\citenamefont{Tanabashi et~al.}(2018)}]{Tanabashi:2018pdg}
\bibinfo{author}{\bibfnamefont{M.}~\bibnamefont{Tanabashi}}
  \bibnamefont{et~al.} (\bibinfo{collaboration}{Particle Data Group}),
  \bibinfo{journal}{Phys. Rev. D 030001} \textbf{\bibinfo{volume}{98}},
  \bibinfo{pages}{030001} (\bibinfo{year}{2018}).

\bibitem[{\citenamefont{Chang et~al.}(2018)\citenamefont{Chang, Essig, and
  McDermott}}]{Chang:2018rso}
\bibinfo{author}{\bibfnamefont{J.~H.} \bibnamefont{Chang}},
  \bibinfo{author}{\bibfnamefont{R.}~\bibnamefont{Essig}}, \bibnamefont{and}
  \bibinfo{author}{\bibfnamefont{S.~D.} \bibnamefont{McDermott}}
  (\bibinfo{year}{2018}), \eprint{1803.00993}.

\bibitem[{\citenamefont{Widrow}(1989{\natexlab{a}})}]{Widrow:1989vj}
\bibinfo{author}{\bibfnamefont{L.~M.} \bibnamefont{Widrow}},
  \bibinfo{journal}{Phys. Rev.} \textbf{\bibinfo{volume}{D40}},
  \bibinfo{pages}{1002} (\bibinfo{year}{1989}{\natexlab{a}}).

\bibitem[{\citenamefont{Widrow}(1989{\natexlab{b}})}]{Widrow:1989fe}
\bibinfo{author}{\bibfnamefont{L.~M.} \bibnamefont{Widrow}},
  \bibinfo{journal}{Phys. Rev.} \textbf{\bibinfo{volume}{D39}},
  \bibinfo{pages}{3576} (\bibinfo{year}{1989}{\natexlab{b}}).

\bibitem[{\citenamefont{Ferrer et~al.}(2018)\citenamefont{Ferrer, Massó,
  Panico, Pujolàs, and Rompineve}}]{Ferrer:2018}
\bibinfo{author}{\bibfnamefont{F.}~\bibnamefont{Ferrer}},
  \bibinfo{author}{\bibfnamefont{E.}~\bibnamefont{Massó}},
  \bibinfo{author}{\bibfnamefont{G.}~\bibnamefont{Panico}},
  \bibinfo{author}{\bibfnamefont{O.}~\bibnamefont{Pujolàs}}, \bibnamefont{and}
  \bibinfo{author}{\bibfnamefont{F.}~\bibnamefont{Rompineve}},
  \bibinfo{journal}{to appear}.

\bibitem[{\citenamefont{Poulin et~al.}(2017)\citenamefont{Poulin, Serpico,
  Calore, Clesse, and Kohri}}]{Poulin:2017bwe}
\bibinfo{author}{\bibfnamefont{V.}~\bibnamefont{Poulin}},
  \bibinfo{author}{\bibfnamefont{P.~D.} \bibnamefont{Serpico}},
  \bibinfo{author}{\bibfnamefont{F.}~\bibnamefont{Calore}},
  \bibinfo{author}{\bibfnamefont{S.}~\bibnamefont{Clesse}}, \bibnamefont{and}
  \bibinfo{author}{\bibfnamefont{K.}~\bibnamefont{Kohri}},
  \bibinfo{journal}{Phys. Rev.} \textbf{\bibinfo{volume}{D96}},
  \bibinfo{pages}{083524} (\bibinfo{year}{2017}), \eprint{1707.04206}.

\bibitem[{\citenamefont{Rai and Senjanovic}(1994)}]{Rai:1992xw}
\bibinfo{author}{\bibfnamefont{B.}~\bibnamefont{Rai}} \bibnamefont{and}
  \bibinfo{author}{\bibfnamefont{G.}~\bibnamefont{Senjanovic}},
  \bibinfo{journal}{Phys. Rev.} \textbf{\bibinfo{volume}{D49}},
  \bibinfo{pages}{2729} (\bibinfo{year}{1994}), \eprint{hep-ph/9301240}.

\bibitem[{\citenamefont{Ringwald and Saikawa}(2016)}]{Ringwald:2015dsf}
\bibinfo{author}{\bibfnamefont{A.}~\bibnamefont{Ringwald}} \bibnamefont{and}
  \bibinfo{author}{\bibfnamefont{K.}~\bibnamefont{Saikawa}},
  \bibinfo{journal}{Phys. Rev.} \textbf{\bibinfo{volume}{D93}},
  \bibinfo{pages}{085031} (\bibinfo{year}{2016}), \bibinfo{note}{[Addendum:
  Phys. Rev.D94,no.4,049908(2016)]}, \eprint{1512.06436}.
  
\bibitem{Kamionkowski:1992mf}
  M.~Kamionkowski and J.~March-Russell,
  Phys.\ Lett.\ B {\bf 282} (1992) 137, 
  \eprint{hep-th/9202003}.
  
\bibitem{Holman:1992us}
  R.~Holman, S.~D.~H.~Hsu, T.~W.~Kephart, E.~W.~Kolb, R.~Watkins and L.~M.~Widrow,
  Phys.\ Lett.\ B {\bf 282} (1992) 132,
  \eprint{hep-ph/9203206}.
  
\bibitem{Barr:1992qq}
  S.~M.~Barr and D.~Seckel,
  Phys.\ Rev.\ D {\bf 46} (1992) 539.


  



\bibitem[{\citenamefont{Kallosh et~al.}(1995)\citenamefont{Kallosh, Linde,
  Linde, and Susskind}}]{Kallosh:1995hi}
\bibinfo{author}{\bibfnamefont{R.}~\bibnamefont{Kallosh}},
  \bibinfo{author}{\bibfnamefont{A.~D.} \bibnamefont{Linde}},
  \bibinfo{author}{\bibfnamefont{D.~A.} \bibnamefont{Linde}}, \bibnamefont{and}
  \bibinfo{author}{\bibfnamefont{L.}~\bibnamefont{Susskind}},
  \bibinfo{journal}{Phys. Rev.} \textbf{\bibinfo{volume}{D52}},
  \bibinfo{pages}{912} (\bibinfo{year}{1995}), \eprint{hep-th/9502069}.
  
\bibitem{Dvali:2005an}
  G.~Dvali,
  \eprint{hep-th/0507215}.


\bibitem[{\citenamefont{Alonso and Urbano}(2017)}]{Alonso:2017avz}
\bibinfo{author}{\bibfnamefont{R.}~\bibnamefont{Alonso}} \bibnamefont{and}
  \bibinfo{author}{\bibfnamefont{A.}~\bibnamefont{Urbano}}
  (\bibinfo{year}{2017}), \eprint{1706.07415}.

\bibitem[{\citenamefont{Hamaguchi et~al.}(2018)\citenamefont{Hamaguchi, Nagata,
  Yanagi, and Zheng}}]{Hamaguchi:2018oqw}
\bibinfo{author}{\bibfnamefont{K.}~\bibnamefont{Hamaguchi}},
  \bibinfo{author}{\bibfnamefont{N.}~\bibnamefont{Nagata}},
  \bibinfo{author}{\bibfnamefont{K.}~\bibnamefont{Yanagi}}, \bibnamefont{and}
  \bibinfo{author}{\bibfnamefont{J.}~\bibnamefont{Zheng}}
  (\bibinfo{year}{2018}), \eprint{1806.07151}.

\bibitem[{\citenamefont{Di~Luzio et~al.}(2017)\citenamefont{Di~Luzio, Mescia,
  Nardi, Panci, and Ziegler}}]{DiLuzio:2017ogq}
\bibinfo{author}{\bibfnamefont{L.}~\bibnamefont{Di~Luzio}},
  \bibinfo{author}{\bibfnamefont{F.}~\bibnamefont{Mescia}},
  \bibinfo{author}{\bibfnamefont{E.}~\bibnamefont{Nardi}},
  \bibinfo{author}{\bibfnamefont{P.}~\bibnamefont{Panci}}, \bibnamefont{and}
  \bibinfo{author}{\bibfnamefont{R.}~\bibnamefont{Ziegler}}
  (\bibinfo{year}{2017}), \eprint{1712.04940}.

\bibitem[{\citenamefont{Giannotti et~al.}(2016)\citenamefont{Giannotti,
  Irastorza, Redondo, and Ringwald}}]{Giannotti:2015kwo}
\bibinfo{author}{\bibfnamefont{M.}~\bibnamefont{Giannotti}},
  \bibinfo{author}{\bibfnamefont{I.}~\bibnamefont{Irastorza}},
  \bibinfo{author}{\bibfnamefont{J.}~\bibnamefont{Redondo}}, \bibnamefont{and}
  \bibinfo{author}{\bibfnamefont{A.}~\bibnamefont{Ringwald}},
  \bibinfo{journal}{JCAP} \textbf{\bibinfo{volume}{1605}}, \bibinfo{pages}{057}
  (\bibinfo{year}{2016}), \eprint{1512.08108}.

\bibitem[{\citenamefont{Giannotti et~al.}(2017)\citenamefont{Giannotti,
  Irastorza, Redondo, Ringwald, and Saikawa}}]{Giannotti:2017hny}
\bibinfo{author}{\bibfnamefont{M.}~\bibnamefont{Giannotti}},
  \bibinfo{author}{\bibfnamefont{I.~G.} \bibnamefont{Irastorza}},
  \bibinfo{author}{\bibfnamefont{J.}~\bibnamefont{Redondo}},
  \bibinfo{author}{\bibfnamefont{A.}~\bibnamefont{Ringwald}}, \bibnamefont{and}
  \bibinfo{author}{\bibfnamefont{K.}~\bibnamefont{Saikawa}},
  \bibinfo{journal}{JCAP} \textbf{\bibinfo{volume}{1710}}, \bibinfo{pages}{010}
  (\bibinfo{year}{2017}), \eprint{1708.02111}.

\bibitem[{\citenamefont{Armengaud et~al.}(2014)}]{Armengaud:2014gea}
\bibinfo{author}{\bibfnamefont{E.}~\bibnamefont{Armengaud}}
  \bibnamefont{et~al.}, \bibinfo{journal}{JINST} \textbf{\bibinfo{volume}{9}},
  \bibinfo{pages}{T05002} (\bibinfo{year}{2014}), \eprint{1401.3233}.

\bibitem[{\citenamefont{Anastassopoulos
  et~al.}(2017)}]{Anastassopoulos:2017kag}
\bibinfo{author}{\bibfnamefont{V.}~\bibnamefont{Anastassopoulos}}
  \bibnamefont{et~al.} (\bibinfo{collaboration}{TASTE}),
  \bibinfo{journal}{JINST} \textbf{\bibinfo{volume}{12}},
  \bibinfo{pages}{P11019} (\bibinfo{year}{2017}), \eprint{1706.09378}.

\bibitem[{\citenamefont{Bähre et~al.}(2013)}]{Bahre:2013ywa}
\bibinfo{author}{\bibfnamefont{R.}~\bibnamefont{Bähre}} \bibnamefont{et~al.},
  \bibinfo{journal}{JINST} \textbf{\bibinfo{volume}{8}},
  \bibinfo{pages}{T09001} (\bibinfo{year}{2013}), \eprint{1302.5647}.

\bibitem[{\citenamefont{Arvanitaki and Geraci}(2014)}]{Arvanitaki:2014dfa}
\bibinfo{author}{\bibfnamefont{A.}~\bibnamefont{Arvanitaki}} \bibnamefont{and}
  \bibinfo{author}{\bibfnamefont{A.~A.} \bibnamefont{Geraci}},
  \bibinfo{journal}{Phys. Rev. Lett.} \textbf{\bibinfo{volume}{113}},
  \bibinfo{pages}{161801} (\bibinfo{year}{2014}), \eprint{1403.1290}.

\bibitem[{\citenamefont{Klein et~al.}(2016)}]{Klein:2015hvg}
\bibinfo{author}{\bibfnamefont{A.}~\bibnamefont{Klein}} \bibnamefont{et~al.},
  \bibinfo{journal}{Phys. Rev.} \textbf{\bibinfo{volume}{D93}},
  \bibinfo{pages}{024003} (\bibinfo{year}{2016}), \eprint{1511.05581}.

\bibitem[{\citenamefont{Saikawa}(2017)}]{Saikawa:2017hiv}
\bibinfo{author}{\bibfnamefont{K.}~\bibnamefont{Saikawa}},
  \bibinfo{journal}{Universe} \textbf{\bibinfo{volume}{3}}, \bibinfo{pages}{40}
  (\bibinfo{year}{2017}), \eprint{1703.02576}.

\bibitem[{\citenamefont{Aasi et~al.}(2015)}]{TheLIGOScientific:2014jea}
\bibinfo{author}{\bibfnamefont{J.}~\bibnamefont{Aasi}} \bibnamefont{et~al.}
  (\bibinfo{collaboration}{LIGO Scientific}), \bibinfo{journal}{Class. Quant.
  Grav.} \textbf{\bibinfo{volume}{32}}, \bibinfo{pages}{074001}
  (\bibinfo{year}{2015}), \eprint{1411.4547}.

\bibitem[{\citenamefont{Punturo et~al.}(2010)}]{Punturo:2010zz}
\bibinfo{author}{\bibfnamefont{M.}~\bibnamefont{Punturo}} \bibnamefont{et~al.},
  \bibinfo{journal}{Class. Quant. Grav.} \textbf{\bibinfo{volume}{27}},
  \bibinfo{pages}{194002} (\bibinfo{year}{2010}).

\bibitem[{\citenamefont{Janssen et~al.}(2015)}]{Janssen:2014dka}
\bibinfo{author}{\bibfnamefont{G.}~\bibnamefont{Janssen}} \bibnamefont{et~al.},
  \bibinfo{journal}{PoS} \textbf{\bibinfo{volume}{AASKA14}},
  \bibinfo{pages}{037} (\bibinfo{year}{2015}), \eprint{1501.00127}.
  
\bibitem{Coleman:1978ae}
S.~R.~Coleman,
 Subnucl. Ser. {\bf 15}, 805 (1979), [382(1978)].
 
\bibitem{Borsanyi:2015cka}
S.~Borsanyi, M. Dierigl, Z. Fodor, S. D. Katz, S. W. Mages,
D. Nogradi, J. Redondo, A. Ringwald, and K. K. Szabo, Phys.
Lett. {\bf B752}, 175 (2016), \eprint{1508.06917}.

\bibitem{Bonati:2016imp}
C. Bonati, M. D’Elia, M. Mariti, G. Martinelli, M. Mesiti,
F. Negro, F. Sanfilippo, and G. Villadoro, EPJ Web Conf. {\bf 137},
08004 (2017), \eprint{1612.06269}.

\end{thebibliography}
\end{document}